\documentclass[desactivate]{aa}   
\newcommand\dscript[2]{\bgroup#1\egroup_\bgroup#2\egroup}
\usepackage{graphicx}
\usepackage{hyperref}
\usepackage{xcolor}
\usepackage{amsmath,amssymb} 
\usepackage[flushleft]{threeparttable}

\newcommand*{\astrosun}{{\odot}}

\newcommand*{\earth}{{\oplus}}

\usepackage{ulem}

\usepackage{txfonts}
\begin{document} 

  \title{Is the high-energy environment of K2-18b special?}
    
    \author{S. Rukdee
          \inst{1} 
          \and
          M. G{\"u}del\inst{2} 
          \and 
          I. Vilović \inst{3}
          \and
          K. Poppenh{\"a}ger\inst{3,4}
          \and
          S. Boro Saikia\inst{2}
          \and
          J. Buchner\inst{1}
           \and \\
          B. Stelzer\inst{5}
          \and
          G. Roccetti\inst{6,7}
          \and
          J. V. Seidel\inst{8,9}
          \and
          V. Burwitz\inst{1}
          }

   \institute{Max Planck Institute for Extraterrestrial     Physics, Giessenbachstrasse 1, 85748 Garching, Germany
         \and
         Department of Astrophysics, University of Vienna, T\"urkenschanzstr. 17, 1180 Vienna, Austria
         \and
         Leibniz-Institute for Astrophysics Potsdam (AIP), An der Sternwarte 16, 14482 Potsdam, Germany
         \and
         Potsdam University, Institute for Physics and Astronomy, Karl-Liebknecht-Str.\ 24/25, 14476 Potsdam-Golm, Germany
         \and
         Institut f{\"u}r Astronomie und Astrophysik, Eberhard Karls Universit{\"a}t T{\"u}bingen, Sand 1, 72076 T{\"u}bingen, Germany
         \and
         European Southern Observatory, Karl-Schwarzschild-Straße 2, 85748, Garching near Munich, Germany
         \and
         Meteorologisches Institut, Ludwig-Maximilians-Universität München, Munich, Germany
         \and
         European Southern Observatory, Alonso de Córdova 3107, Vitacura, Región Metropolitana, Chile
         \and
         Laboratoire Lagrange, Observatoire de la Côte d’Azur, CNRS, Université Côte d’Azur, Nice, France
         }

   \date{Received 8 October 2025; accepted 12 January 2026}

 
  \abstract
   {K2-18~b lies near the radius valley that separates super-Earths and sub-Neptunes, marking a key transitional regime in planetary and atmospheric composition. The system offers a valuable opportunity to study how M-dwarf high-energy stellar radiation influences atmospheric stability and the potential for sustaining volatile species, especially important in the context of the upcoming ELT and its ANDES spectrograph.}
   {This study characterizes the high-energy environment of K2-18 with X-ray observations from eROSITA, the soft X-ray instrument on the Spectrum-Roentgen-Gamma (SRG) mission, Chandra, and XMM-Newton.}
   {We derive a representative 0.2-2 keV X-ray flux with an APEC thermal plasma model fitted with the Bayesian X-ray Analysis (BXA). With the observed X-ray flux from the exoplanet host star, we estimate the photoevaporative mass loss of exoplanet K2-18b using the energy-limited model. In addition, we examine the thermal structure of the system based on a hydrodynamic model.
   }
   {In 100 ks XMM-Newton observation we identified K2-18 as a very faint X-ray source with $\mathrm{F_X = 10^{-15}\ erg\ s^{-1}\ cm^{-2}}$, with an activity level of (Lx/Lbol) $\sim 10^{-5}$. A small flare has been detected during the observation. The planet is irradiated by an X-ray flux of $\mathrm{F_{pl,X} = 12\pm3\ erg\,s^{-1}\,cm^{-2}}$.}
  {The X-ray flux measurement of K2-18 gives important limitations for atmospheric escape and photochemical modeling of its exoplanets. Despite its near orbit around an M-dwarf star, K2-18b's low activity level environment suggests that it can retain an atmosphere, supporting recent tentative detections of atmospheres.}
   \titlerunning{Is the high-energy environment of K2-18b special?}
   \authorrunning{S. Rukdee et al.}
   \maketitle

%


\section{Introduction}

High-energy radiation from an exoplanet's host star can cause a planet's atmospheric evaporation \citep{Lammer_2003, Vidal-Madjar2003, Ehrenreich2011, Johnstone2018, Yan2018, Poppenhaeger2021, Foster2022, Fromont2024}. Understanding the host star's X-ray and Ultraviolet (XUV) flux, combined with modeling, helps determine its impact on atmospheric escape, chemistry, and long-term atmospheric evolution \citep{Becker_2020, Johnstone2021, Amaral_2025, McCreery_2025, VanLooveren2025}. Many small planets discovered to date are around low-mass M-type stars \citep{Howard_2012, Dressing_2013}. The long pre-main-sequence lifetime of M stars implies that the star remains strongly magnetically active for much longer than a G star \citep{Ramirez2014}. Close proximity of exoplanets to their host stars therefore increases the risk of high-energy radiation affecting their physical and chemical properties. Multiple studies identify XUV radiation from M dwarfs as a potential threat to planetary habitability \citep{Heath1999, Tarter2007, Lammer2009, Shields2016, Meadows2018, VanLooveren2024, VanLooveren2025}. M-dwarfs are known for their frequent and intense flares \citep[e.g.,][]{Loyd_2018, Gunther_2020}. Strong flares from an M-dwarf host star can significantly alter the chemistry of a planet's atmosphere and may lead to its removal \citep{Gudel2002, Segura2010}. Early James Webb Space Telescope (JWST) results suggest that temperate rocky exoplanets around M dwarfs do not retain substantial atmospheres \citep{Greene2023, Zieba2023, Cadieux2024}.
These observations have meanwhile been explained using atmospheric mass-loss models \citep{VanLooveren2025}. For instance, the Sun produces its most energetic flares ($\sim\mathrm{10^{32}\ erg}$) about once per solar cycle \citep{Lin1994, Aulanier2013, youngblood2017}, whereas M~dwarfs can emit flares of similar energy daily \citep{Audard2000}. Even flares from magnetically inactive early M-type stars, which exhibit fewer starspots and less frequent flaring, can significantly influence the atmospheric chemistry of orbiting planets \citep{Hawley2014}. Previous studies show that M dwarfs exhibit lower X-ray luminosities compared to G stars at any age, but they take longer to desaturate in terms of Lx/Lbol, especially late M dwarfs \citep{Magaudda2020, Johnstone2021_stars}. A recent study \citep{Zhu2025} of neighboring GKM stars finds that the majority ($\geq$60\%) of nearby M-dwarfs no later than M6 have X-ray activity ($\mathrm{\log(L_X/L_{bol})}$) levels that are not higher than the average ($\mathrm{80^{th}}$ percentile) recorded in G-type stars. The considerable diversity of stars and their exoplanets motivates detailed characterization of individual systems.

Differences between UV and X-ray emissions in the high-energy environment indicate various physical processes occurring in different layers of the stellar atmosphere. UV radiation is largely emitted by the chromosphere and transition region of stars, where temperatures range from thousands to tens of thousands of Kelvin.  Magnetic activity and ionization processes heat these locations \citep{France2016, Ribas_2009}. The corona, the outermost and hottest layer of the stellar environment, generates X-rays at temperatures of millions of Kelvin. X-rays are produced by very energetic processes such as magnetic reconnection, which leads to plasma heating \citep{Shoda2021,Maggio_2023}. Accurate observations across the UV and X-ray bands are critical for understanding the physical mechanisms that drive star emissions.

The wavelength-dependent photoabsorption cross-sections of atoms and molecules cause diverse heating and chemical processes across the planetary atmospheric layers. In high-energy environments, NUV (1800-3200 $\AA$), FUV (912-1800 $\AA$), and X-ray (5-100 $\AA$) photons are absorbed in the middle to upper atmosphere, photo-dissociating molecules and ionizing heavy elements.  EUV photons (100-911 $\AA$) are absorbed higher in the thermosphere, ionizing atoms and molecules, leading to both thermal and non-thermal atmospheric escape \citep{France2016, Youngblood2016, Gronoff2020}. Lyman-$\alpha$ (Ly$\alpha$), commonly used as a proxy for high-energy UV flux, accounts for approximately 37\%–75\% of the total 1150–3100 $\AA$ flux in most M dwarfs \citep{France2013}.

K2-18 hosts an exoplanetary system \citep{Montet_2015} and is an excellent target for atmospheric studies. K2-18 is a M2.8 star with a rotation period of $\sim$38.6 days \citep{Cloutier2017}. Of particular interest is the planet K2-18b, which orbits the star with an orbital period of 33 days at a semi-major axis of approximately 0.142 AU \citep{Montet_2015}. A wide range of interior structure models have been proposed for K2-18b, involving different proportions of $\mathrm{H_2O}$ and $\mathrm{H_2}$/He in a volatile-rich atmosphere. These scenarios were followed up by observations \citep{Madhusudhan2023} revealing $\sim$1\% $\mathrm{CO_2}$, $\sim$1\% $\mathrm{CH_4}$, and non-detections of CO and $\mathrm{NH_3}$ as shown in Figure 1 in \cite{Luu_2024}. Similar abundances of $\mathrm{CO_2}$ and $\mathrm{CH_4}$ were also reported by \cite{Hu2025}. Proposed scenarios include a thin-atmosphere rocky world \citep{Tsai2021, Yu2021}, a thin-atmosphere water world sometimes referred to as a 'Hycean' world \citep{PietteMadhusudhan2020, Madhusudhan2021, Madhusudhan2023, Cooke2024, Tsai2024, Wogan2024}, a deep-atmosphere mini-Neptune lacking a solid surface \citep{Hu2021, Tsai2021, Yu2021, Wogan2024}, and a magma ocean mini-Neptune \citep{Shorttle2024}. A rocky world scenario appears inconsistent with current observations \citep{Madhusudhan2020}, primarily due to the planet’s bulk density ($\mathrm{2.67^\dscript{+0.52}{-0.47} g/cm^2}$ from \cite{Benneke2019}) and atmospheric composition. In contrast, the thin atmosphere-water world and mini-Neptune scenarios remain plausible. Nonetheless, the Hycean scenario itself is questioned by the community, even under the assumptions of the abundances derived in \citet{Madhusudhan2023} \citep[e.g.][]{Wogan2024, Huang2024, Shorttle2024, Werlen2025}.

Among these interior studies, the Stratified Mini-Neptune model \citep{Benneke2024} proposes a homogeneous supercritical $\mathrm{H_2}$–$\mathrm{H_2O}$ ocean at pressures between 1–5\,kbar, consistent with the observed $\mathrm{CH_4}$/$\mathrm{CO_2}$ ratio. The Phase-Separated Mini-Neptune Scenario \citep{Gupta2025} suggests that pressures of 5-10 kbar, $\mathrm{H_2}$ and $\mathrm{H_2O}$ may be immiscible at some temperatures, resulting in a phase-separated water-rich layer underneath a supercritical $\mathrm{H_2}$-rich layer. Additionally, \citet{Luu_2024} proposes a global hydrothermal system where a supercritical water ocean interacts with the atmosphere. This model explains the observed $\mathrm{CH_4}$/$\mathrm{CO_2}$ ratio and the non-detection of CO, suggesting ocean temperatures ranging from 710 K to 1070 K. According to the decision tree roadmap for characterizing temperate sub-Neptunes from \citet{Hu2025}, K2-18b could fall into the category of a mixed steam envelope with nitrogen depletion. 

For the atmospheric observation on K2-18b, previously, \cite{Benneke2019} and \cite{Santos2020} suggested that K2-18b possesses an H$_2$-dominated envelope with potential HST detection of $\mathrm{H_2O}$ \citep{Tsiaras2019}. The EUV irradiation on K2-18b was estimated to lie in the range of $\mathrm{10^1 - 10^2~erg~s^{-1}~cm^{-2}}$ from \cite{Santos2020}. They find that under this EUV it is likely to lose only a small fraction of its mass (1\% or less) over its remaining lifetime. The study suggested that the planet is probably not an archetypal planet crossing the radius valley \citep{Fulton2017, Fulton_Petigura2018, VanEylen2018} to become a bare rock. 

Recently, from the JWST near-infrared data, a tentative detection of the potential biomarkers dimethyl sulfide (DMS), and dimethyl disulfide (DMDS) was reported \citep{Madhusudhan2025}. On Earth, DMS is mainly produced by marine phytoplankton through biological processes \citep{Madhusudhan2025}. While these claimed detections of such complex molecules are unique in exoplanet studies—where focus is typically placed on simpler species like $\mathrm{H_2O}$, $\mathrm{CO_2}$, and $\mathrm{CH_4}$—it’s important to note that the reported DMS claim in \citet{Madhusudhan2025} stems from a limited retrieval framework. As highlighted by \citet{Welbanks_2025}, the molecular input set in the original DMS detection was highly constrained, which significantly influenced the inferred composition and contributed to the ongoing debate over the potential biosignature claim. A re-analysis of the original detection data by \citet{Schmidt2025} found both the claimed detection of $\mathrm{CO_2}$ and DMS unsubstantiated, while they were able to reproduce the $\mathrm{CH_4}$ detection. With independent analysis to uncover statistically significant spectral features, \cite{Taylor2025}, \cite{Welbanks_2025} and \cite{Stevenson2025} find no significant feature in the same JWST data. Most importantly, \cite{Luque_2025} jointly analyzed NIRISS, NIRSpec, and MIRI data over the full panchromatic spectrum and found insufficient evidence for DMS or DMDS and that various molecules with methyl functional groups provide an equally good fit to the data. While the biomarker detection claim on K2-18~b has, therefore, been robustly discussed and subsequently been dismissed by the community \citep{Pica_Ciamarra_2025,Luque_2025, Welbanks_2025, Taylor2025}, the debate on the detectability of potential biomarkers in exoplanets orbiting M-dwarfs remains a main science driver in our field. In this context, understanding how K2-18 compares with other M-dwarfs in terms of X-ray output is essential for assessing the long-term habitability of its planets, in particular K2-18b, and their suitability for more targeted observational campaigns. Indeed, a tentative distinction of DMS from a flat line fit would require $\sim$25 more MIRI transits \citep{Luque_2025} and only a holistic understanding of K2-18b's space environment could potentially justify the extraordinary time commitment. Previous research has highlighted the potential for atmospheric characterization of K2-18b; however, direct constraints on its high-energy environment are limited due to the scarcity and low sensitivity of previous X-ray observations. Understanding the planet's atmospheric composition and chemistry is highly dependent on a precise understanding of the incident stellar flux, which has remained particularly unknown in the XUV regime.

High-resolution spectroscopic studies \citep{Sairam2025}  of chromospheric lines (H$\alpha$, CaII H \& K) show that K2-18 exhibits relatively low chromospheric activity, consistent with its rotation period and photometric variability, which indicated low activity during recent JWST observations. This quiet state is favorable for atmospheric characterization of K2-18b, as it minimizes contamination in transmission spectra and stellar variability. However, residual chromospheric activity may still contribute to the high-energy radiation environment. Long-term photometric monitoring further suggests the presence of an activity cycle \citep{Sairam2025}, providing useful constraints for planning future spectroscopic follow-up of this system.

In this work, we revisit the high-energy environment of the star K2-18 to determine how long K2-18b can retain its atmosphere. Even if the atmosphere is not fully lost, its chemistry and haze formation are shaped by the stellar high-energy radiation, making it critical to study this environment for insights into both atmospheric escape and photochemical processes. By presenting new measurements of K2-18's X-ray flux, we aim to refine current models of atmospheric evolution and improve our understanding of habitability around low-mass stars. This is especially relevant in the context of future telescopes designed to evaluate the habitability of rocky exoplanets orbiting nearby M-dwarfs, such as through reflected light observations with ANDES on the ELT \citep{Palle2025, Roccetti2025}. In Sect.~2, we present new X-ray observations from different missions over the past four years. In Sect.~3, we describe the spectral extraction, spectral analysis, and modeling methods used to interpret the data. The results of the recent X-ray observations are shown in Sect.~4. We discuss these results in Sect.~5 and conclude in Sect.~6.

\subsection*{Star K2-18}
The star K2-18 is located approximately $\mathrm{38.07\pm0.08}$ pc from the Sun. It is classified as an M2.5V-type star \citep{Schweitzer2019}, with a mass of $\mathrm{0.413 \pm 0.043,M_\astrosun}$, a radius of $\mathrm{0.394\pm 0.038,R_\astrosun}$, effective temperature of $\mathrm{3503 \pm 60}$ K and a metallicity of $\mathrm{0.09 \pm 0.09}$ dex \citep{Montet2015}. \cite{Hejazi2024} used high-resolution spectroscopic data from IGRINS to derive abundance ratios of planet-building elements, including Al/Mg, Ca/Mg, Fe/Mg, and C/O. The C/O yields $\mathrm{0.568 \pm 0.026}$. Later, \cite{Sairam2025} used archival photometric and spectroscopic observations to estimate the stars' effective temperature ($T_{\mathrm{eff}}$) of 3645$\pm$52 K  and age of 2.9–3.1 Gyr. Their metallicity estimates ($\mathrm{[Fe/H] = 0.10\pm0.12\, dex}$) are in good agreement with previous studies. \cite{Guinan_2019} adopted the age of 2.4 Gyr for K2-18 and use $F_{\rm X}$-Age and $\mathrm{F_{Ly\alpha}}$-Age ratios to estimate X-ray and Ly$\alpha$ (FUV) irradiances on K2-18b to be $\mathrm{F_x \sim 29\pm8 \,erg\,s^{-1}\,cm^{-2}}$ and  $\mathrm{F_{\rm Ly\alpha} \sim 61\pm20 \,erg\,s^{-1}\,cm^{-2}}$. The estimated X-ray and Ly$\alpha$ irradiances of K2-18b are approximately 115× and 8× higher than Earth's \citep{Guinan_2019}. This indicates extensive atmospheric change through photo-evaporation (see also \citealt{VanLooveren2025}, Figs. 3 and 5), especially during the early evolution when the X-ray luminosity must have been even higher given the star's higher rotation rate (\citealt{Johnstone2021_stars}, hereafter CJ+21). In addition, \cite{Santos2020} derived the high energy environment from HST $Ly\alpha$ observation to be $\mathrm{F_{Ly\alpha} \sim 100.7^\dscript{+96.1}{-82.4} \,erg\,s^{-1}\,cm^{-2}}$ and estimated the EUV flux of $\mathrm{107.9^\dscript{+124.7}{-90.8} \,erg\,s^{-1}\,cm^{-2}}$ on the K2-18b following formulae for estimating EUV fluxes from \cite{Linsky2014}. 

\subsection*{Exoplanets around K2-18}
\begin{figure}[h!]
    \centering
 \includegraphics[width=0.5\textwidth]{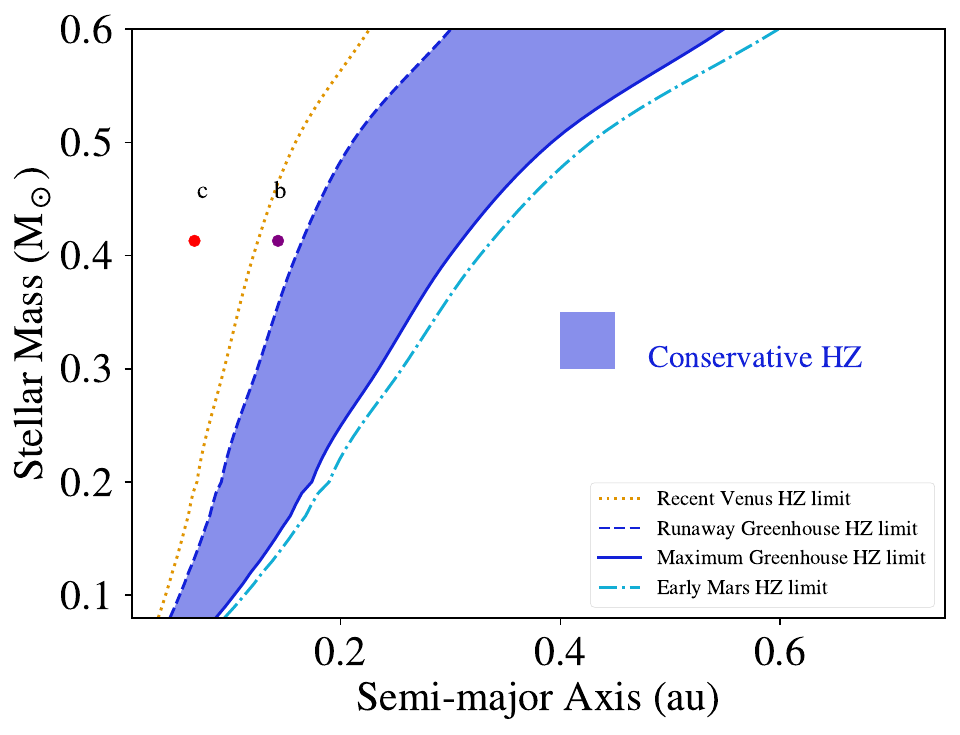}
    \caption{Orbital positions of the K2-18 planets (colored circles) are shown relative to the habitable zone (HZ) \citep{Kopparapu2013}. K2-18b lies near the inner edge of the conservative HZ. The yellow dotted line marks the Recent Venus limit, while the region between the cyan dash-dotted line and the conservative HZ corresponds to orbital distances where water would likely remain frozen. The white regions beyond all HZ boundaries represent environments where surface liquid water is not expected to persist.}
    \label{fig:HZ}
\end{figure}

Two transiting objects around K2-18 were initially observed with the Kepler telescope as part of the K2 mission \citep{Montet2015} as summarized in Table \ref{tab:companions}. Later, \cite{Benneke2017} confirmed the planetary nature of the transit signal by detecting the same transit depth at a different wavelength (4.5 $\mu$m) using the Spitzer Space Telescope. The mass of K2-18b was independently determined by \cite{Cloutier2017} using the HARPS spectrograph and by \cite{Sarkis2018} with the CARMENES spectrograph. \cite{Cloutier2017} determined the mass of the K2-18b to be $\mathrm{8.0\pm1.9 M_\earth}$. Later, the second planet K2-18c was confirmed with both radial velocity instruments \citep{Cloutier2019} as a non-transiting planet of minimum mass $\mathrm{5.62\pm0.84 M_{\earth} \sin i}$. The K2-18b with an orbital period of 33 days has an equilibrium temperature of $\mathrm{272\pm15 K}$ \citep{Montet2015}. Figure \ref{fig:HZ} shows the orbital positions of the planets relative to the habitable zone (HZ), defined according to \cite{Kopparapu2013}, where surface liquid water could exist. \cite{Kopparapu2013} describes the Recent Venus limit as an empirical inner boundary of the optimistic habitable zone. This is based on the possibility that Venus may have had liquid water until not long ago in its history. If an exoplanet gets no more energy from its star than Venus did during that period, it could still have conditions where surface water remains stable. A runaway greenhouse effect occurs when greenhouse gases such as $\mathrm{H_2O},\mathrm{CH_4},\mathrm{CO_2},\mathrm{N_2O}, \mathrm{O_3}$ trap outgoing thermal radiation, leading to rapid heating and the potential loss of surface liquid water. If water vapor reaches the stratosphere, it may escape via hydrodynamic escape, ultimately desiccating the planet \citep{Nakajima1992}. K2-18b is between the Recent Venus habitable zone (HZ) limit and the runaway greenhouse HZ limit, while K2-18c is closer to the host star.

\begin{table}[ht!]
 \centering
    \caption{The exoplanet companions of K2-18}
    \begin{tabular}{cccccc}
    \hline\hline
         Planet & $M_P$ &  $R_P$ &  a & $P $ \\
        & [$M_{\earth}$] & [$R_{\earth}$] &  [AU] & [days] \\ 
         \hline
          c$^c$ & $5.62\pm0.84$* & non-transiting & 0.067 & 8.96 \\ 
        b$^b$ & $8.0\pm1.9$  & 2.38 & 0.143 & 32.9 \\ 
    \hline
    \end{tabular}
    \begin{tablenotes}
    \small
      \item  {Reference $^b$ \cite{Cloutier2017}; $^c$ \cite{Cloutier2019} *minimum mass ($m\sin i$) }
    \end{tablenotes}
    \label{tab:companions}
\end{table}

\section{Observations} \label{sec:observation}
The observational data used in this high-energy environment study came from eROSITA as well as from the archives of Chandra and XMM-Newton, where K2-18 was detected adjacent to the quasar QSO B1127+078.

\subsection{eROSITA}
eROSITA (extended ROentgen Survey with an Imaging Telescope Array) is a wide-field X-ray telescope on board the Russian-German Spectrum-Roentgen-Gamma (SRG) observatory \citep{Predehl2021}. The eROSITA observation has provided only an upper limit for this source. The data has been processed by the eSASS pipeline version 020 (Brunner et al. 2022). We report the eRASS:1 upper limit from the DR1 public release data using the upper-limit server \citep{Tubin2024, Merloni_2024}. 

\subsection{Chandra}
K2-18 lies in archival Chandra data in an ACIS-S observation targeting SDSSJ113017.37+073212.9	(PI Piconcelli, ObsID 25339). Within a total of 4 ks in December 2023, there was no detection.

\subsection{XMM-Newton}
\begin{figure*}[h]
    \centering
    \includegraphics[width=\textwidth]{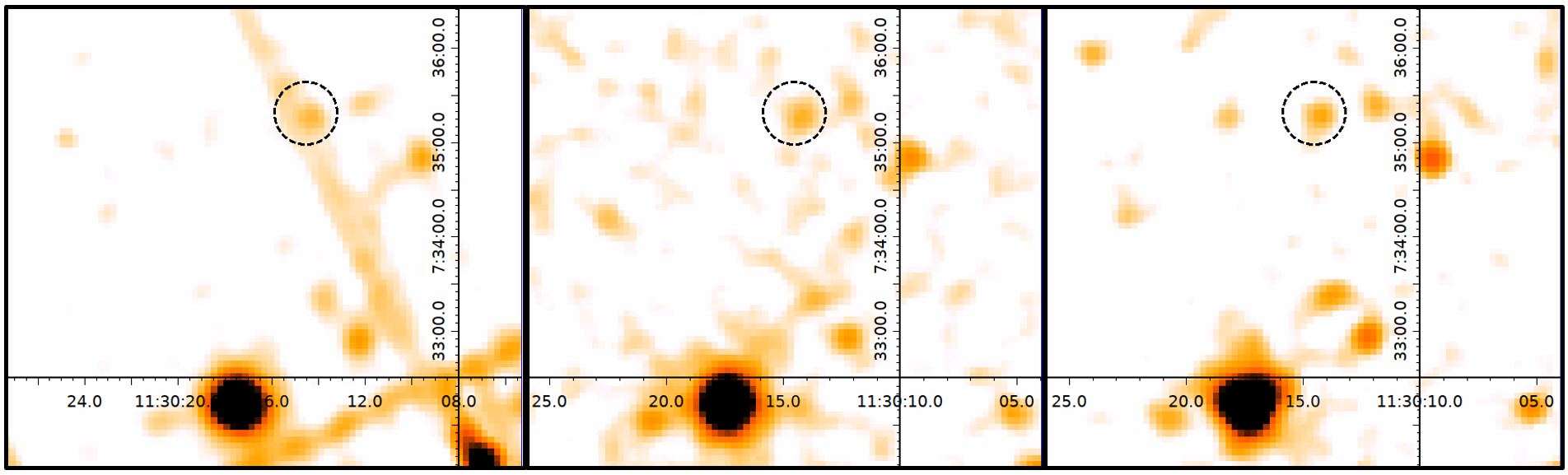}
    \caption{The K2-18 source detected on PN-CCD, MOS1, and MOS2 is marked by the black dashed circle region of 20 arcsec with its coordinates, while the dark region below corresponds to a quasar. The image is displayed with a smoothing parameter of 1.5$\sigma$. Note that the efficiency/sensitivity of MOS2 is lower than MOS1 \citep{Mineo2024}. In PN, the source lies in a chip gap, while it is unobstructed in MOS1 \& MOS2. The X-ray flux was derived in the same way for both MOS and PN detectors.
    }
    \label{fig:XMMimage}
\end{figure*}

K2-18 was detected during a Heritage program observation (PI: Giorgio Lanzuisi) of the quasar WISSH39 (J1130+0732). The dataset (ObsID 0943530501) includes 3 EPIC, 24 OM, and 2 RGS exposures, totaling 110 ks of observation time in December 2023. For K2-18, the data from the PN, MOS1, and MOS2 cameras are shown in Fig.\ref{fig:XMMimage}. Each detector has a single exposure of 110 ks.

\begin{figure*}[h!]
    \centering
 \includegraphics[width=0.33\textwidth]{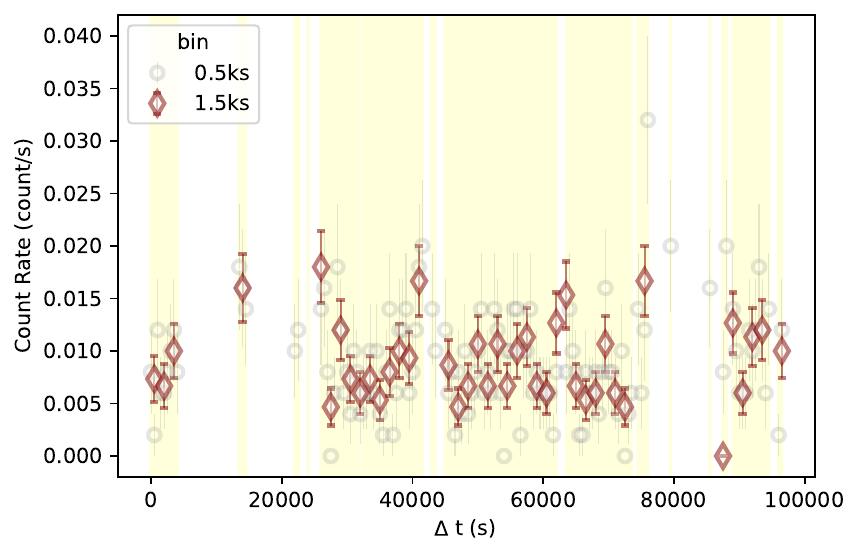}
 \includegraphics[width=0.33\textwidth]{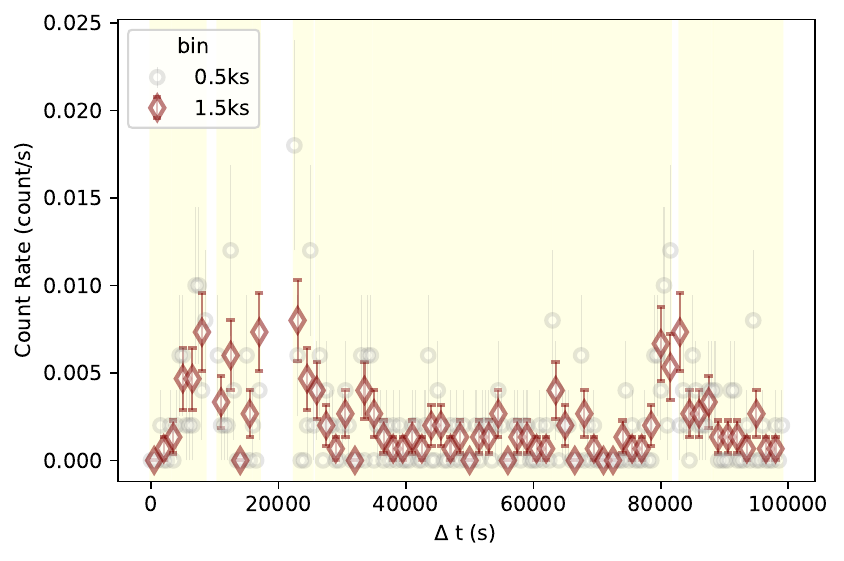}
 \includegraphics[width=0.33\textwidth]{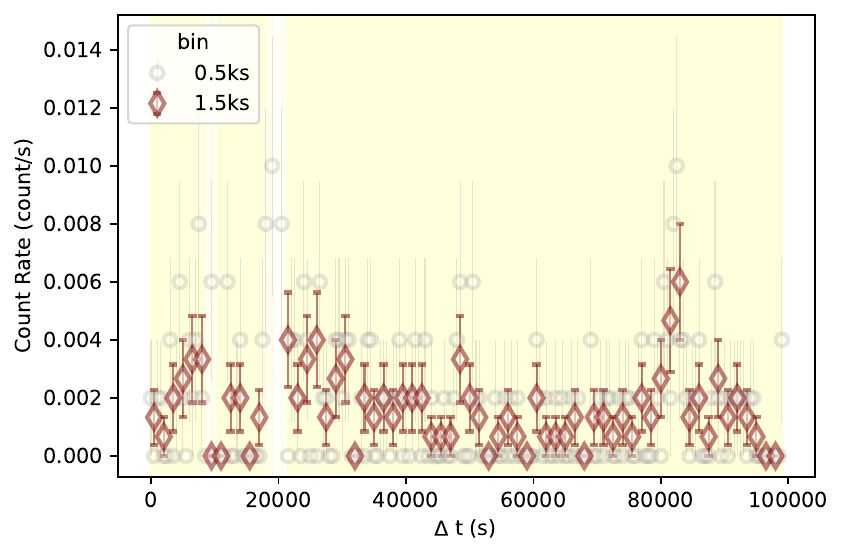}

    \caption{K2-18 XMM-Newton light curves from left to right: PN, MOS1, and MOS2, showing quasi-quiescent activity. The yellow background marks the good time interval (GTI) from each detector. Two light curves are plotted with different time binning (grey 0.5 ks bin and brown 1.5 ks bin).} 
    \label{fig:XMM_lc}
\end{figure*}

\section{Methods} \label{sec:methods} 
In this section, we describe source extraction and spectral fitting method in Sect~\ref{subsec:spectral_analysis}. The model for X-ray driven evaporation of the planet atmospheres is in Sect \ref{subsec:planet_evo}.

\subsection{Spectral Analysis} \label{subsec:spectral_analysis}
In this work, we perform two independent spectral analyses: (1) a Bayesian inference approach, and (2) the Hardness Ratio method. The data reduction was carried out using standard pipelines, specifically CIAO/Sherpa for Chandra data and XMM/SAS for XMM-Newton observations.

\subsubsection{Spectral extraction}

The Chandra data were analyzed using standard procedures from the CIAO package version 4.14 \citep{Fruscione2006}. We filtered the event list to the 0.3–10 keV energy range. Source counts were extracted using circular regions with a radius of 5", while the background was estimated from an annular region surrounding the source. The background region has an inner radius of 10" and an outer radius of 20". From the Chandra observation, we place an upper limit on our analysis. We proceed consistently with the detected sources, and analyze the source spectrum and background region spectra. The sampled posterior contains a wide range of normalizations, allowing also very faint source solutions that give no photons. As with the other spectra, we compute the posterior distribution of fluxes from the spectral parameter posterior distribution. We adopt the (1$\sigma$) upper error bar of the fluxes as a flux upper limit for Chandra.

We processed the XMM-Newton PN, MOS1, and MOS2 data (see Fig. \ref{fig:XMMimage}) following standard procedures in the XMM-Newton User Handbook. To filter out background noise\footnote{\url{https://www.cosmos.esa.int/web/xmm-newton/sas-thread-epic-filterbackground-in-python}}, we used specific selection rules for each of the EPIC detectors. We extract a high-energy light curve using only single-pixel events (PATTERN==0) to identify intervals affected by flaring particle background. For EPIC-MOS, the selection uses $\mathrm{(PI>10000)}$, whereas for EPIC-pn it is restricted to  $\mathrm{(10000<PI<12000)}$, with the upper bound applied to avoid misidentification of hot pixels as very high-energy events. Good time intervals (GTIs) were generated by applying a rate threshold (RATE $\leq$ 0.4 counts/s for PN and RATE $\leq$ 0.35 counts/s for MOS) with \texttt{tabgtigen}, and filtered event files were created accordingly. The difference in thresholds reflects the distinct instrumental responses of MOS and PN, with PN generally exhibiting higher background levels. These GTIs are used to filter the event list, ensuring that only low-background data are retained for further analysis. The data were barycenter-corrected using \texttt{barycen}. Light curves were extracted from a 20" circular region for the source and an annular region (60" inner radius and 120" outer radius) for the background using \texttt{evselect}, with energy filtering applied in the 0.2–10 keV range. Light curves were corrected for instrumental effects with \texttt{epiclccorr}. Following data reprocessing, we examined the light curves in the datasets from each XMM-Newton detector. The light curves showed a quasi-quiescent level of stellar activity. Figure~\ref{fig:XMM_lc} displays the XMM-Newton light curve showing a potential flare towards the end of the observation. As K2-18 is located at a close distance to the solar system, we did not include a model component for ISM absorption, since it is negligible for stellar distances below 100\,pc. 

As our source shows relatively low counts, it is important to extract as much information as possible from the available data. This is best done by modelling the spectral contribution of the background carefully in addition to the source, and analysing with Poisson statistics (C-Stat). For the choice of the spectral model, there are two approaches we consider. In maximum likelihood analyses, one may use as simple a spectral model as justified by the data. For simplicity, we employ such an approach below, with a single-temperature plasma from the Astrophysical Plasma Emission Code \texttt{APEC} model \citep{Smith2001}. 

An alternative is Bayesian inference, which allows marginalizing over possible spectral models, without needing to restrict the model based on the data quality. We describe this approach here, presented in detail in \cite{Rukdee2024}. To make Bayesian inference with X-ray spectral analysis computationally feasible, we use the spectral fitting package Bayesian X-ray Analysis, \texttt{BXA} \citep{Buchner2014}. \texttt{BXA} integrates the nested sampling algorithm \texttt{UltraNest} \citep{Buchner2021} with the fitting environment \texttt{CIAO/Sherpa} \citep{Fruscione2006}. To balance computational efficiency with physical accuracy, we constrain all metal abundances to a single value and model the plasma using ten APEC components on a logarithmic temperature grid. Normalizations follow a Gaussian profile, approximating a continuous temperature distribution. This approach captures the plasma behavior more accurately than single- or two-temperature models while remaining computationally tractable within the \texttt{BXA} framework\footnote{\url{https://github.com/SurangkhanaRukdee/BXA-Plasma}}. Free parameters include peak temperature, normalization, and log-Gaussian distribution width $\sigma$. Table~\ref{tab:prior} provides a full list of priors that control the normalization of each component. In addition, the abundance parameter sets the metal abundances of all \texttt{APEC} model components. This also includes inferred Differential
Emission Measure (DEM) distributions, which quantify the amount of plasma emitting X-rays at a given temperature, and is derived by fitting the relative contributions of plasma at different temperatures to the observed spectrum. We adopt the temperature distribution outlined in \cite{Rukdee2024} to characterize the plasma behavior.

For the background region spectrum, we adopt the background model of \cite{Simmonds2018} obtained through principal components analysis (PCA), and fit this model to each background spectrum with BXA. Throughout our analysis, we employ Poisson (C-stat) statistics and jointly fit the source and background spectra. The background model is not a physical model, but an empirical one that is fitted to the background spectrum. The background spectral fit parameters are then fixed during the analysis of the source spectrum (which includes both source and background). The empirical background model spectrum parameters are the normalizations of fixed PCA components. However, the addition is performed in log space (to ensure non-negative count rates). The PCA components for XMM-Newton were trained on background spectra from the XMM-Newton archive \citep{Simmonds2018}. Essentially, this approach approximates the extracted background spectrum with a smooth function, with a preference for creating background spectral models that resemble other XMM-Newton backgrounds. The background model is shown in the result in Sect. 4.1.

\begin{table}[h]
    \centering
    \caption{Priors set for the APEC model}
    \begin{tabular}{llc}
    \hline\hline
        Parameter & Prior & Range\\
        \hline
        abundance & uniform & 0.0 - 1.0\\
        $kT_{\rm peak}$ & log-uniform & 0.1 - 5.0\\
        norm$_{\rm peak}$ & log-uniform & $10^{-6}$ - 0.01\\
        $\sigma$ & uniform & 0.0 - 2.0\\
    \hline
    \end{tabular}
    \label{tab:prior}
\end{table}

We independently confirmed the results of the Bayesian analysis that we present in Sect.~\ref{sec:results} by a more traditional spectral analysis with Xspec for the extracted CCD spectra from MOS1 and MOS2, i.e.\ the detectors where the source was not located on a chip edge, with similar outcomes for the coronal temperature and stellar X-ray fluxes. Furthermore, we confirmed the overall coronal temperature reported in Sect.~\ref{sec:results} with a hardness ratio analysis along the lines of \cite{Ilic2022}, i.e.\ calculating the hardness ratio of the star with 0.2-0.7~keV as the soft band and 0.7-2.0~keV as the hard band and comparing this to expected hardness ratios for single-temperature coronal models, again with comparable results to the outcomes of the Bayesian analysis. 

The upper limit of the background flux was computed in the 0.6–2.3 keV band for Chandra's while the X-ray flux is computed in the 0.2–2.0 keV band for XMM-Newton. For the flux conversion, we convert the HST Ly$\alpha$ to EUV  flux using the method from \cite{Linsky2014} for an M2.5 star, applying the conversion factors listed in Table~\ref{tab:lya} and summing over all bands from 10 to 117~nm. For the EUV-to-X-ray relation, we adopt the  \cite{Sanz_Forcada2011} (hereafter SF+11) relation. In SF+11 the EUV flux was calculated by generating a synthetic spectrum using a coronal model based on the Emission Measure Distribution (EMD), coronal abundances, and the Astrophysics Plasma Emission Database (APED; \citealt{Smith2001}). 

\begin{table}[h]
\centering
\caption{Relations for estimating EUV flux from Ly$\alpha$ for M dwarfs (M2.5 spectral type) from \cite{Linsky2014}}
\begin{tabular}{c c}
    \hline\hline
Wavelength Band (nm) & $\log \left[\frac{f(\Delta \lambda)}{f(\mathrm{Ly}\alpha)}\right]$ \\
    \hline
10--20    & $-0.491$ \\
20--30    & $-0.548$ \\
30--40    & $-0.602$ \\
40--50    & $-2.294 + 0.258 \log[f(\mathrm{Ly}\alpha)]$ \\
50--60    & $-2.098 + 0.572 \log[f(\mathrm{Ly}\alpha)]$ \\
60--70    & $-1.920 + 0.240 \log[f(\mathrm{Ly}\alpha)]$ \\
70--80    & $-1.894 + 0.518 \log[f(\mathrm{Ly}\alpha)]$ \\
80--91.2  & $-1.811 + 0.764 \log[f(\mathrm{Ly}\alpha)]$ \\
91.2--117 & $-1.004 + 0.065 \log[f(\mathrm{Ly}\alpha)]$ \\
    \hline
     \label{tab:lya}
\end{tabular}
\end{table}

\subsection{Modeling Stellar and Planetary Evolution} \label{subsec:planet_evo}
For the stellar evolutionary model linked to atmospheric escape, we explore an approach using the hydrodynamic model with energy-limited escape, implemented via the \texttt{VPLanet} framework \citep{Barnes2020}. The results are also put in context with Jeans' escape calculations of Earth's early evolution \citep{Johnstone2021}. These models provide context for our discussion and help interpret the observational data. \citet{Salz2016} and \citet{Krenn2021} performed a critical assessment of the applicability of the energy-limited approximation \citep{Watson1981} for estimating exoplanetary mass-loss rates by comparing it to a grid of hydrodynamic models computed by \citet{Kubyshkina_2018}. They concluded that the energy-limited approximation can be used as an order-of-magnitude estimate for planets with intermediate gravitational potentials and low-to-intermediate equilibrium temperatures (300 K - 2000 K) and XUV irradiation levels. However, the XUV irradiation level depends on the evolution of the stellar rotation period of the host star, which is non-unique for a 0.4 solar mass star up to 2 Gyr \citep{VanLooveren2025}. We note that the work of \cite{VanLooveren2025} focuses on atmospheres dominated by $\mathrm{CO_2}$ and $\mathrm{N_2}$, whereas K2-18b is currently believed to host an $\mathrm{H_2}$-rich atmosphere. Furthermore, the planet's higher mass enhances atmospheric retention by making it more difficult for gases to escape compared to a lower-mass planet.

We use the Model for Rotation of Stars (\texttt{MORS})\footnote{\url{https://github.com/ColinPhilipJohnstone/Mors}} from CJ+21 to describe the stellar evolution. It provides a comprehensive rotation–$L_{\rm X}$–EUV–age framework spanning the full main-sequence evolution of cool stars, including non-unique pathways before 1 Gyr. It models different scenarios of evolutionary tracks depending on the rotation rate percentile, with the three scenarios being 'slow', 'medium', and 'fast', corresponding to the 5th, 50th, and 95th percentiles of the rotation distribution. For this model, \citet{Johnstone2021} examines the minimum $\mathrm{CO_2}$ levels required to retain an atmosphere under high XUV flux, focusing on young stars and early Earth-like conditions. Later, \citet{VanLooveren2025} introduced the Atmosphere Retention Distance (ARD) for Earth-sized planets in stellar habitable zones, combining thermochemical and stellar evolution models. Their analysis accounts for stellar rotation rates and associated XUV output to assess where $\mathrm{CO_2}$- or $\mathrm{N_2}$-dominated atmospheres can be retained across a range of stellar masses. 

\subsubsection*{Energy-limited mass loss} 
Next, we investigate the host star's role in planetary mass loss. We determined the mass loss rate with the energy-limited model \citep{Lopez2012, Owen2012} using the following equation:
\begin{equation}
\dot{M} = \epsilon\times \frac{\pi R_{XUV}^2 F_{XUV}}{KGM_{pl}/R_{pl}}
\label{eq:1}
\end{equation}
where $\dot{M}$ is the mass loss rate, $\epsilon$ the efficiency factor from atmospheric escape assumed to be 0.15 \citep{Foster2022}, $R_{XUV}$ the planetary radius in XUV wavelength adopted to be 1.1 times the planetary radius in optical, $M_{pl}$ the planetary mass, $K$ is a factor indicating the effect of Roche-lobe overflow and set to 1, $G$ the gravitational constant, and $F_{\rm XUV}$ the sum of the observed X-ray and estimated EUV fluxes of the host star. Here, we derive the EUV luminosity from the X-ray luminosity following Eq.~3 in SF+11. 

The efficiency factor $\epsilon$ measures the fraction of incoming stellar XUV energy that is converted into work to drive atmospheric escape. For hydrogen–helium atmospheres, $\epsilon$ is usually modeled on the order of 10 to 30\% (see e.g. \citealt{Lammer_2013}, \citealt{owen_wu2013}, \citealt{Salz2016}, \citealt{Poppenhaeger2025}). This range is based on early hydrodynamic simulations which showed that a significant fraction of the absorbed energy is radiated away since much of the absorbed energy is lost to radiative cooling, mainly through Lyman-$\alpha$ emission \citep{Salz2015}. These values, however, come almost entirely from hydrogen-dominated models \citep{Watson1981}, in which heavier species such as C, N, and O are not expected to escape. For atmospheres enriched in heavier molecules, a large share of the XUV input may instead drive photochemistry, leaving even less energy available for escape. This makes $\epsilon$ highly uncertain for mixed or metal-rich envelopes like that of K2-18b. Even for pure H/He atmospheres, hydrodynamic simulations \citep{Krenn2021} show that energy-limited estimates do not reliably match mass-loss rates in any regime, and the uncertainties remain large. For these reasons, we use the energy-limited expression only as a qualitative diagnostic and under the assumption of a hydrogen-only atmosphere, with $\epsilon$ remaining a major and unresolved source of uncertainty in atmospheric escape estimates for sub-Neptune planets. When applying the energy-limited formula, we adopt a conservative efficiency factor (0.15). This is within the range well supported by detailed simulations in the literature \citep{Shematovich2014}.

We calculate the XUV flux received by the planet ($F_{\rm pl,xuv}$) by dividing the total stellar XUV luminosity ($L_{\rm tot,xuv}$) by the surface area of a sphere with a radius equal to the planet's orbital distance ($a$), $\mathrm{F_{\rm pl} = L_{\rm total} / (4\pi a^2)}$. The resulting flux is expressed in units of erg s$^{-1}$ cm$^{-2}$.

In addition to Eq.~\ref{eq:1}, we use the \texttt{VPLanet} suite to provide an overview of planetary evolution. Due to limitations in the mini-Neptune radius model embedded in \texttt{VPLanet}, we describe the method and present the results in the Appendix~\ref{sec:VPLanet-results}.
\section{Results} \label{sec:results}

In this section, we present the results from the analysis of X-ray observations obtained from Chandra and XMM-Newton. These observations provided X-ray properties and the activity levels of K2-18. Additionally, we report the upper limit value from eROSITA observations according to eROSITA Data Release 1 from the upper limit server \citep{Tubin2024, Merloni_2024}. 

\subsection{X-ray properties from spectral analysis} 
\label{sec:xrayproperties}
\begin{figure}
    \centering
    \includegraphics[width=0.5\textwidth]{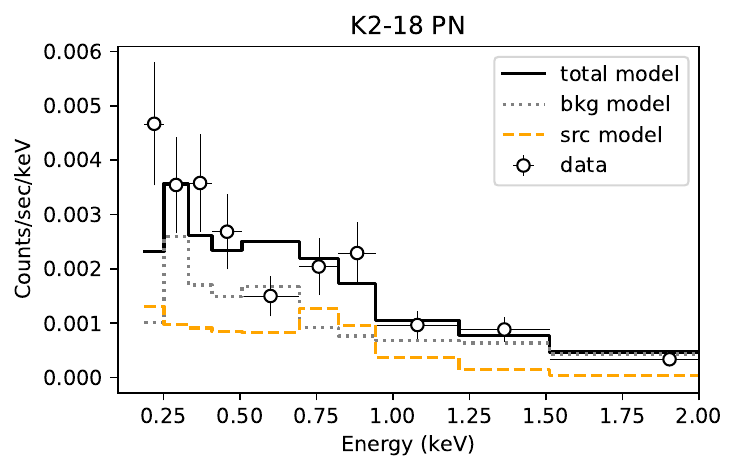}  
    \includegraphics[width=0.5\textwidth]{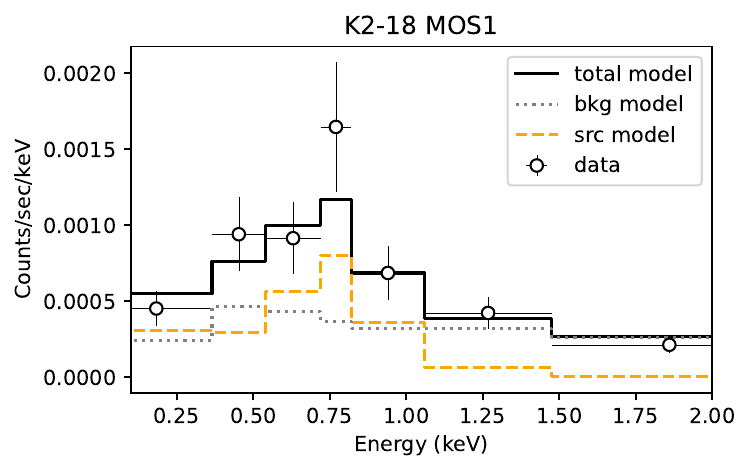} \includegraphics[width=0.5\textwidth]{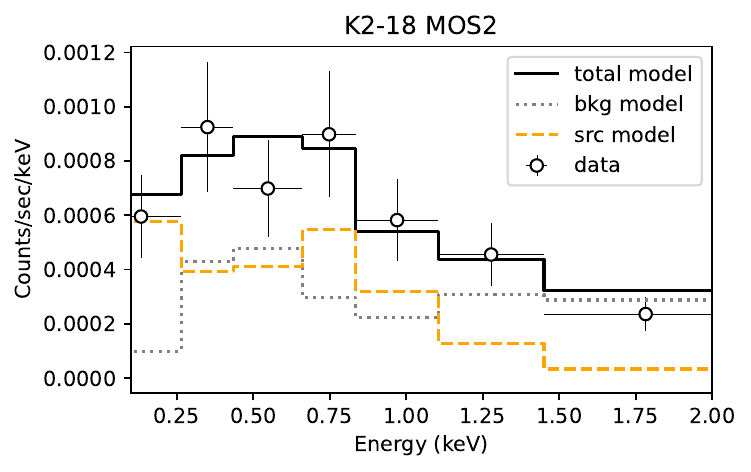}
    \caption{ X-ray spectrum of K2-18 analyzed over the 0.2–2.0 keV range, extracted from the PN-CCD (top), MOS1 (middle), and MOS2 (bottom) on XMM-Newton during a 100 ks observation. The black solid line is the spectral model through \texttt{BXA-Plasma}, the orange dashed line is the source model, and the grey dotted line is the background model with PCA routine in \texttt{BXA}.}
    \label{fig:EPIC_spec}
\end{figure}

\begin{figure}
    \centering
    \includegraphics[width=0.35\textwidth]{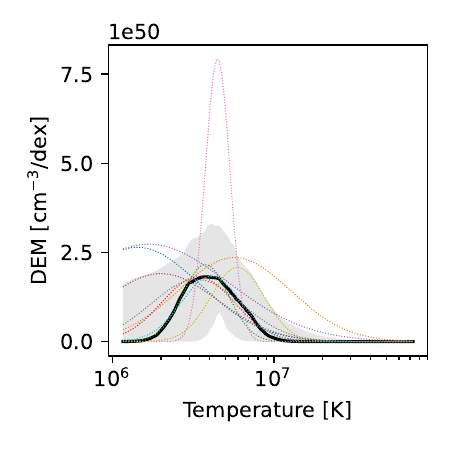}
    \caption{ The reconstructed continuous Differential Emission Measure (DEM) distribution from MOS1. The black curve shows the median of the posterior DEM predictions, and the gray band contains 68\% of the distribution. The colorful dotted lines are ten random posterior samples of possible DEM distributions. Note that the kT values from PN and MOS2 are not well constrained and are not shown here.} 
    \label{fig:spec_temp}
\end{figure}

\begin{table*}
    \begin{center}
    \caption{Soft X-ray flux properties for K2-18 across different observations
    \label{tab:properties}}
    \begin{tabular}{ccccccccc}
    \hline\hline
         Date & Observatory  & Duration & Flux & $\mathrm{logLx}$ & $\mathrm{L_x/L_{bol}}$ & $kT_{peak}$  & $kT_{\sigma}$ & Abundance\\
          &   & [ks] & [$\mathrm{10^{-15}\, erg\,cm^{-2}\,s^{-1}}$] & [erg/s] &[$\mathrm{10^{-5}}$] &[keV] & [keV] & [Solar] \\
         \hline

          2020 & eROSITA &  6.20 & $< 140.73$  & $< 28.38$ & & -  & - & - \\
          2023-12 & Chandra &  40.0 & $< 9.09$ &  $< 27.20$  &  & -& - & - \\
          2024-12 & XMM PN* &  110.0 & $5.89^\dscript{+1.04}{-0.96}$   & $27.01\pm0.08$ & $1.33\pm0.23$ & $0.51\pm0.72$  & $0.86\pm0.48$  & $0.15\pm0.12$\\  
          2024-12 & XMM MOS1 &  110.0 & $4.13^\dscript{+1.12}{-0.98}$  & $26.85\pm0.11$ & $0.93\pm0.24$ & $0.35\pm0.15$  & $0.28\pm0.26$  & $0.46\pm0.24$ \\  
          2024-12 & XMM MOS2 &  110.0 & $4.36^\dscript{+0.97}{-0.97}$  &  $26.88\pm0.10$ & $0.98\pm0.22$ & $0.42\pm0.50$  & $0.63\pm0.42$  & $0.18\pm0.17$ \\   
      
    \hline
    \end{tabular}
    \end{center}
    \tiny{*Flux calculated for energy band 0.2-2.0 keV for XMM Newton}
\end{table*}

The spectral fits and parameter constraints from \texttt{BXA} for the XMM-Newton data, with a good time interval (GTI) duration of 57166.05 s, 91064.42, and 95408.17 for PN, MOS,1 and MOS2 respectively, are shown in Fig.\ref{fig:EPIC_spec}. 

The inferred Differential Emission Measure (DEM) distributions, computed with \texttt{APEC}, are presented in Fig.~\ref{fig:spec_temp} for the reconstructed continuous DEM, with the average distribution shown as a black solid line. For the XMM-Newton observation, the DEM is poorly constrained due to the low counts. We obtained the best statistical constraints for the DEM from MOS1, which is shown in Fig.~\ref{fig:spec_temp}. The peak temperature ($kT_{\mathrm{peak}}$) and the width of the distribution ($kT_{\sigma}$) from all detectors are listed in Table \ref{tab:properties}. The abundance was not tightly constrained, but approximately 0.15 solar in PN and MOS2, which is not ruled out by the constraints from MOS1. The focus of this work is however on the X-ray flux, which the low counts confidently constrain to a notably low value. We calculated the X-ray flux of $4.13^\dscript{+1.12}{-0.98}\times10^{-15} \mathrm{erg\,s^{-1}\,cm^2}$ from MOS1. We obtained comparable X-ray fluxes of $5.89^\dscript{+1.04}{-0.96}$ $\mathrm{erg\,s^{-1}\,cm^2}$ and $4.36^\dscript{+0.97}{-0.97}$ $\mathrm{erg\,s^{-1}\,cm^2}$  for PN and MOS2, respectively. The differences between the PN and MOS results are minimal, on the order of $\sim 0.15$ dex in $\log L_{\mathrm{X}}$. From the Chandra observation, we placed a 1$\sigma$ upper limit of $\mathrm{<9.09\times10^{-15} erg\,s^{-1}\,cm^2}$. 

\begin{figure}[h!]
    \centering
    \includegraphics[width=0.5\textwidth]{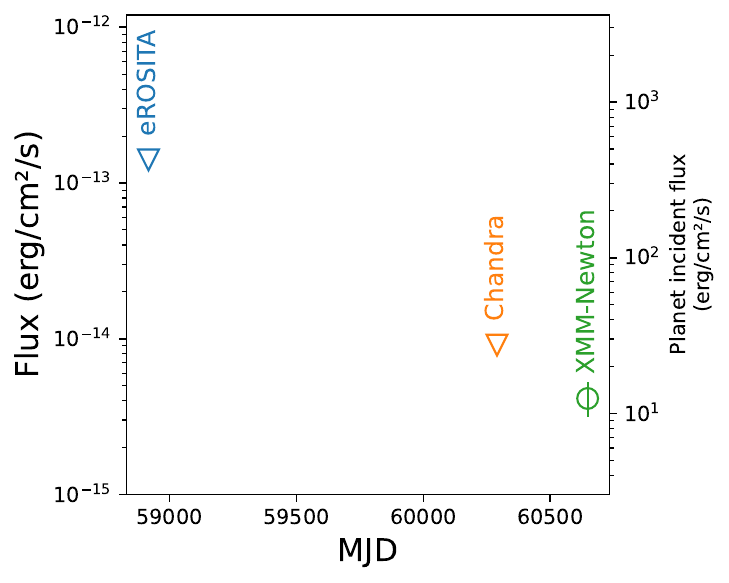}
    \caption{The light curve of the high energy X-ray flux over four years comprises upper limit from eROSITA and Chandra, and XMM-Newton measurement.}
    \label{fig:xrayobs}
\end{figure}

\subsection{X-ray and UV context}

The light curve of the X-ray flux over time from various missions over approximately four years is presented in Figure~\ref{fig:xrayobs}. 
Before the X-ray observations, the star K2-18 was observed during two transits on June 18, 2017, and March 9, 2018 (Program GO-14221, PI: D. Ehrenreich) using HST/STIS and the grating G140M (resolving power R$\sim$10,000). The HST Ly$\alpha$ flux was taken from Figure 3 of \citet{Santos2020}, using the average value of data points to represent the stellar environment. We selected the red wing of the Ly$\alpha$ line, before the transit, which appears more stable and better constrained, and find a Ly$\alpha$ flux of $0.78^\dscript{+0.19}{-0.16}\times\mathrm{10^{-15}\, erg\,cm^{-2}\,s^{-1}}$ for the first visit (Visit A) and $1.81^\dscript{+0.23}{-0.23}\times\mathrm{10^{-15}\, erg\,cm^{-2}\,s^{-1}}$ for the second visit (Visit B). Ly$\alpha$  flux serves as a proxy for the total FUV flux, accounting for approximately 75\% to 90\% of the FUV emission in M dwarfs \cite{Guinan_2019}. We did not include the EUV flux converted to X-ray in Fig.\ref{fig:xrayobs} as it was found to be unrealistically low.

In this work, the XMM-Newton detection provides a tight and direct constraint on the X-ray flux. Results from our Bayesian method (Fig. \ref{fig:EPIC_spec} and Fig.~\ref{fig:spec_temp}) are consistent with a flux of $3.70^{+0.74}_{-0.69}\times 10^{-15} \mathrm{erg\,s^{-1}\,cm^{-2}}$ derived from the simpler Hardness Ratio method, corresponding to an X-ray luminosity $\sim 7\times10^{26}~\mathrm{erg\,s^{-1}}$. The ratio ($\mathrm{L_{XUV}/L_{bol} \sim 10^{-5}}$), indicates a low stellar activity level at the present day. 

We estimated the stellar EUV luminosity ($\mathrm{L_{EUV}}$) from Ly-$\alpha$ flux from HST visit B reported by \cite{Santos2020} using the \cite{Linsky2014} relation (range 10-117 nm) to be $2.7 \times 10^{26}$ erg s$^{-1}$ and combined with our X-ray measurement resulting in XUV luminosity of  $9.78 \times 10^{26}$ erg s$^{-1}$. The corresponding XUV flux received by K2-18b is $\sim17.03\, \mathrm{erg\,s^{-1}\,cm^{-2}}$. 

For an instantanous estimation based on energy-limited escape of the atmosphere, we calculate an atmospheric mass-loss rate driven by XUV radiation from Eq.\ref{eq:1} to be $1.07 \times 10^7$ g s$^{-1}$, which is an order of magnitude less than the total escape rate of $10^8$ g s$^{-1}$ previously reported by \citet{Santos2020} from the HST measurement. This is further illustrated in the planetary evolution model from VPLanet \citep{Barnes2020}, shown in Fig.~\ref{fig:atm_escape}, where the planetary mass and radius exhibit minimal change over a billion-year timescale. 

\begin{figure}[h!]
    \centering
    \includegraphics[width=0.5\textwidth]{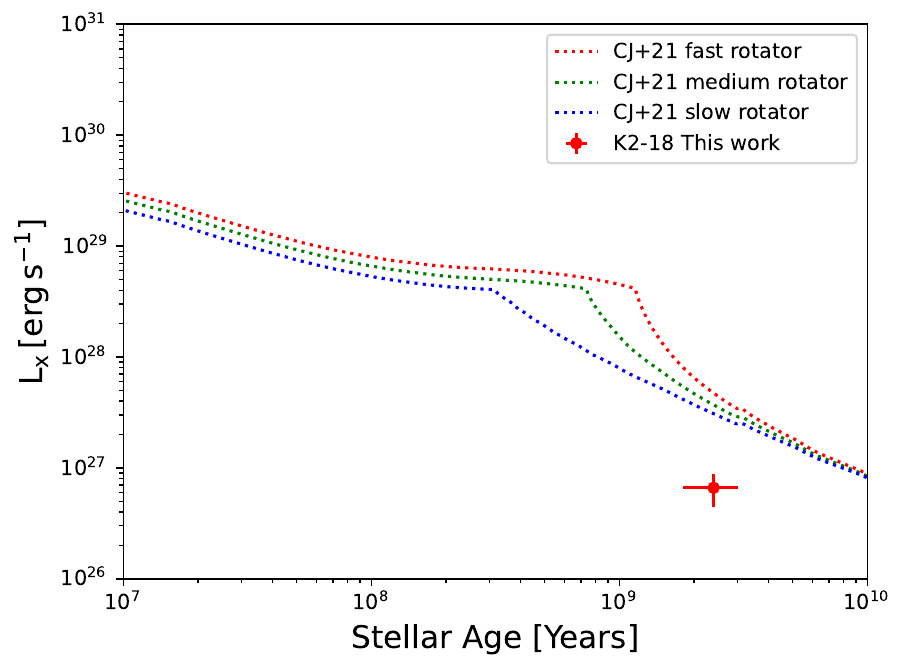}
        \caption{The stellar evolutionary track of X-ray luminosity from \citet{Johnstone2021_stars} (CJ+21). The age of K2-18 is adopted from \cite{Guinan_2019}. The dotted lines show the evolution track generated with \texttt{MORS} for different percentiles of the rotation distribution: (red) fast rotator, (green) medium rotator and (blue) slow rotator. The measured X-ray luminosity is almost an order of magnitude lower than the expectation from the model.}
    \label{fig:evolutionary_track}
\end{figure}

The measured value from the MOS1 observation is placed onto Fig.~\ref{fig:evolutionary_track}, which shows the stellar evolution track from CJ+21 appropriate for K2-18 ($\mathrm{M_*}$ = 0.41$\mathrm{M_\astrosun}$, Prot = 38.6 days at present). We find that the \cite{Lehmer_2017} model of mini-Neptune has a maximum planet radius limitation at 2.2 $\mathrm{R_{\oplus}}$ shown in Fig. \ref{fig:atm_escape}. Alternatively, we do not apply the planet model and instead leave the radius fixed at the present-day measured planet radius of 2.38  $\mathrm{R_{\oplus}}$. The resulting envelope mass-loss rate is $6.2\times 10^{-7} \ \mathrm{M_{\oplus} \ Myr^{-1}}$ from \texttt{VPLanet}, whereas with the energy limited mass loss calculation (Eq.\ref{eq:1}) using the present-day flux yields $5.63\times 10^{-8} \ \mathrm{M_{\oplus} \ Myr^{-1}}$ which is about an order of magnitude lower than what the model predicted. Here we again see a discrepancy in the mass-loss rate that may be related to the similarly large discrepancy between the modeled and the observed X-ray luminosities in Fig. \ref{fig:evolutionary_track}.

\section{Discussion} \label{sec:discussion}

Is the observed level of XUV flux sufficiently low to allow atmospheric retention? Current observations suggest that K2-18b possesses an atmosphere. Interior structure models indicate that planets in this size range likely have a solid or rocky core enveloped by a thick, hydrogen-rich atmosphere \citep{Cloutier2017}, with volatiles such as water potentially surrounding the core \citep{Tsiaras2019}. 

In this work, we presented the observed X-ray flux of the system K2-18.  In the context of stellar activity, the surface X-ray flux of low-mass stars, particularly M dwarfs, spans a wide range, reflecting their magnetic activity levels \citep{Caramazza2023}. Low X-ray emission of such stars is often consistent with other indicators—such as low metallicity and slow rotation, suggesting an old age and a quiescent magnetic environment. Unlike previous assumptions of a constant $L_X$ level in the saturated regime, \citet{Magaudda2020} show that $L_X$ decreases slightly with increasing $P_{\mathrm{rot}}$, that the saturated $L_X$ level decreases with lower stellar mass, and that the transition point ($P_{\mathrm{rot, sat}}$) shifts to longer periods for lower-mass stars. The $L_X$–$P_{\mathrm{rot}}$ relation for K2-18’s stellar mass and rotation period places the system in the unsaturated regime, with an expected $\log L_X \sim 27$ erg\,s$^{-1}$.

\begin{figure}[h!]
    \centering
   \includegraphics[width=0.5\textwidth]{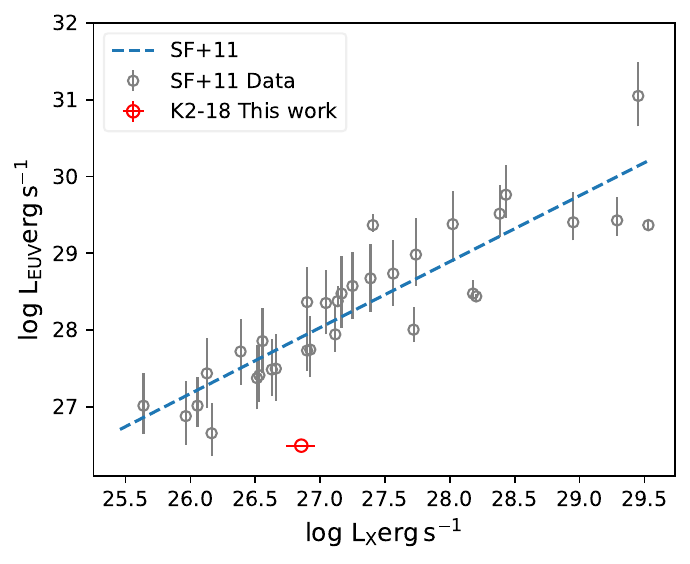}
    \caption{ 
    X-ray and UV luminosity of the K2-18 system, with Ly$\alpha$ emission representing the EUV flux and XMM-Newton observations. The X-ray flux plotted against the empirical relation derived from sample (grey data points) from SF+11.The red point is given by its observed XMM-Newton X-ray luminosity and by the EUV luminosity inferrred from  with Ly$\alpha$ from \cite{Santos2020}.}    \label{fig:SF11}
\end{figure}

Previous observations by \citet{Santos2020} using Ly$\alpha$ transit spectroscopy provided evidence for hydrogen escape from K2-18b’s atmosphere, likely driven by EUV radiation. Their study estimated an EUV irradiation level of $10^1$–$10^2 \,\mathrm{erg\,s^{-1}\,cm^{-2}}$ using the energy-limited escape from \cite{Salz2015} corresponding to a loss of less than 1\% of the planet’s mass over its lifetime, while allowing it to maintain a volatile-rich atmosphere. \cite{Santos2020} concluded that K2-18b possesses an H-rich atmosphere. We included the Ly$\alpha$ flux measurement from \citet{Santos2020}, converted to EUV using the relation from \citet{Linsky2014}, along with the X-ray measurement from this work, and found that K2-18 lies an order of magnitude below the SF+11 relation. This is shown in Fig.~\ref{fig:SF11}. With an energy-limited escape model over the evolution of the star, we also find that K2-18b can retain its atmosphere (see Appendix~\ref{sec:VPLanet-results}).

Previous observations by \cite{Montet2015} estimate the equilibrium temperature of K2-18b -- derived from the star's bolometric luminosity -- to be similar to Earth’s average surface temperature of approximately 288~K. According to the present-day measured rotation and stellar mass, K2-18 should have fallen onto the fast rotator track. However, the measured X-ray luminosity of the system is lower than expected for a $\mathrm{0.4M_{\astrosun}}$ star with a rotation period of 38.6 days. Previous findings from ground-based observations suggest that low-activity M dwarfs are more common among older, less massive stars with reduced magnetic activity \citep{Houdebine2012, Robertson2013, Maldonado2017}.

\begin{figure}[h!]
    \centering
    \includegraphics[width=0.45\textwidth]{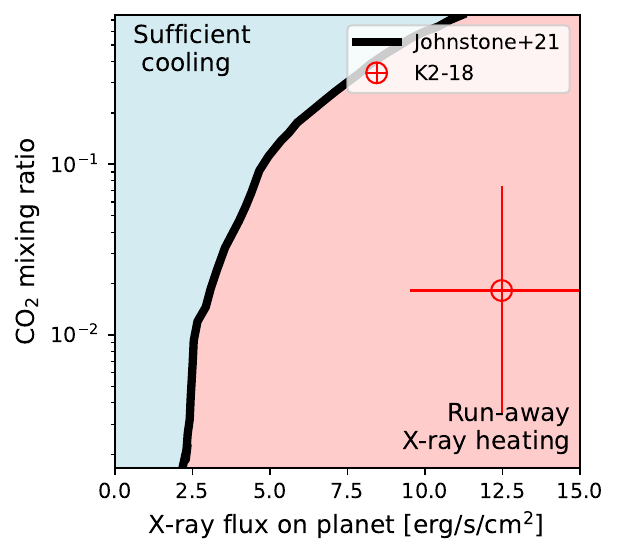}
        \caption{Thermal structure relation of $\mathrm{CO_2}$ (cooling process) and X-ray flux (heating process) on the planet, adapted from \citet{Johnstone2021}. The overlaid boundary (black solid line) corresponds to a $\mathrm{CO_2}$- and $\mathrm{N_2}$-dominated atmosphere, while K2-18b is expected to be H/He-rich with $\mathrm{CO_2}$ as a minor species. The X-ray flux is measured in this work, and the $\mathrm{CO_2}$ mixing ratio is taken from \citet{Madhusudhan2023} (no offset).} 
    \label{fig:thermal_structure}
\end{figure}

The incident flux on the planet and its effect on the temperature profile or thermal structure have been studied using various models. Intense XUV radiation leads to heating of the upper atmosphere, causing it to expand and form extended exospheres \citep{Lammer2007}, while carbon-bearing species such as $\mathrm{CO_2}$ affect the cooling process \citep{Johnstone2021}. General circulation models (GCMs) show that stellar irradiation also influences large-scale atmospheric dynamics, including circulation patterns, cloud formation, and observable properties such as phase curves and albedo. For super-Earths with rocky cores and secondary atmospheres, retention depends on factors such as atmospheric composition, stellar mass, and stellar evolution. \cite{VanLooveren2025} demonstrated that for a $\sim$1 Earth-mass planet, the host star's XUV flux and activity history play a key role in determining long-term atmospheric survival.

Figure \ref{fig:HZ} shows that the planet resides between the Recent Venus and Runaway Greenhouse habitable zone limits. Observations showing the presence of $\mathrm{CH_4}$ in K2-18b’s atmosphere possibly indicate an ongoing runaway greenhouse effect, which is strengthened by the thermal structure context in Fig.\ref{fig:thermal_structure}.  Here, we place observational constraints on the model from \cite{Johnstone2021}, where the measurement of the $\mathrm{CO_2}$ mixing ratio of $10^{(-1.75^\dscript{+0.45}{-1.03})}$ is taken from the no-offset results of \cite{Madhusudhan2023}. The no-offset case refers to the baseline scenario where the NIRISS and NIRSpec spectra are combined directly, without introducing any relative flux adjustments between the instruments or channels. However, if K2-18b is hydrogen-rich, then $\mathrm{H_3^+}$ would play a more significant role in cooling the atmosphere \citep{Miller2010}, while $\mathrm{CO_2}$ would have little influence on the cooling process. The X-ray flux on the planet of $\mathrm{F_{pl,x} = 12.47^{+3.38}_{-2.96} \ erg \ s^{-1} \ cm^{-2}}$ is from XMM-Newton MOS1 (this work). In Fig.\ref{fig:thermal_structure}, the overlaid model curve of \cite{Johnstone2021} describes the balance between cooling, driven by carbon-bearing species such as $\mathrm{CO_2}$ and heating from the high-energy flux incident on the planet. At face value, our X-ray measurement (red error bar in Fig.\ref{fig:thermal_structure}) implies that there is not enough $\mathrm{CO_2}$ to prevent run-away X-ray heating in K2-18b. We note that this model assumes a $\mathrm{CO_2}$- and $\mathrm{N_2}$-dominated atmosphere, whereas K2-18b is expected to be H/He-rich with $\mathrm{CO_2}$ only as a minor species. In addition, \citet{Turbet2019} showed that planets undergoing runaway greenhouse evolution develop thick, water-vapor-dominated atmospheres, significantly inflating their radii—an effect that is observable from space missions. For a better understanding of diverse exoplanetary thermal structures, further observations of sub-Neptunes are essential to place stronger constraints on this model.

Our team, under the MARSH (Methane Atmospheres Related to Stellar Hosts) collaboration, is conducting observations with XMM-Newton and VLT/CRIRES+ to constrain thermal structure models, following the frameworks of \citet{Johnstone2021} and \citet{VanLooveren2025}. These efforts aim to better understand the relevant conditions by populating the parameter space shown in Fig.~\ref{fig:thermal_structure} and refining the models for different stellar types, which should also be considered. Looking ahead, future instruments such as ANDES on the Extremely Large Telescope (ELT) aim to characterize non-transiting temperate rocky exoplanets around nearby M dwarfs, including Proxima Centauri b and Barnard’s Star b. These targets form a "golden sample" \citep{Palle2025} of nearby potentially habitable worlds accessible for detailed atmospheric studies.

\section{Conclusions} \label{sec:conclusion}
This study investigates, from a stellar perspective, why K2-18b appears capable of retaining its atmosphere despite orbiting an M-dwarf star with close proximity to its host. The high-energy radiation environment of exoplanets plays a crucial role in shaping atmospheric escape, chemical evolution, and potential habitability. By combining X-ray and EUV observations with atmospheric spectroscopy, we can begin to map star–planet interactions and understand how stellar activity influences exoplanetary atmospheres over time. We find that K2-18 is relatively inactive as evidenced by its low X-ray luminosity, although small flares are observed. Its activity level (Lx/Lbol) is an order of magnitude higher than that of the present-day Sun. This activity range (Lx/Lbol is 0.93-1.33 $\times10^{-5}$) of the host star may be a sweet spot for future atmospheric characterization-- active enough to drive detectable atmospheric signals, yet not so extreme as to cause rapid atmospheric erosion.

K2-18's X-ray flux measurement provides essential constraints for atmospheric escape and planetary photochemical modeling. The host stars of the most promising habitable planets may be similarly quiet. To accurately characterize the spectra of such quiet stars, reasonable exposure times and instruments with higher effective areas are required. The upcoming NewAthena \citep{Cruise2025} mission, with its increased telescope effective area together with the WFI camera \citep{Rau_2013} will be essential for advancing studies of exoplanetary high-energy environments. 

\begin{acknowledgements}
SR thanks Vinay Kashyap and Nick Durham for a discussion regarding the Chandra calibration. KP and IV acknowledge support from the European Research Council (ERC) under grant agreement 101170037 (Evaporator). JVS and GR acknowledge support by the Munich Institute for Astro-, Particle and BioPhysics (MIAPbP) which is funded by the Deutsche Forschungsgemeinschaft (DFG, German Research Foundation) under Germany´s Excellence Strategy – EXC-2094 – 390783311. We thank the anonymous referee for their constructive comments, which have helped improve this manuscript.

This work is based on data from eROSITA, the soft X-ray instrument aboard SRG, a joint Russian-German science mission supported by the Russian Space Agency (Roskosmos), in the interests of the Russian Academy of Sciences represented by its Space Research Institute (IKI), and the Deutsches Zentrum für Luft- und Raumfahrt (DLR). The SRG spacecraft was built by Lavochkin Association (NPOL) and its subcontractors, and is operated by NPOL with support from the Max Planck Institute for Extraterrestrial Physics (MPE). 

The development and construction of the eROSITA X-ray instrument was led by MPE, with contributions from the Dr. Karl Remeis Observatory Bamberg \& ECAP (FAU Erlangen-Nuernberg), the University of Hamburg Observatory, the Leibniz Institute for Astrophysics Potsdam (AIP), and the Institute for Astronomy and Astrophysics of the University of Tübingen, with the support of DLR and the Max Planck Society. The Argelander Institute for Astronomy of the University of Bonn and the Ludwig Maximilians Universität Munich also participated in the science preparation for eROSITA. 

The eROSITA data shown here were processed using the eSASS/NRTA software system developed by the German eROSITA consortium.
\end{acknowledgements}
\bibliography{ref}
\bibliographystyle{aa}

%
%

%


\begin{appendix} 
\section{\texttt{VPLanet} model} 
\label{sec:VPLanet-results}
We use the \texttt{VPLanet} model \citep{Barnes2020} to evaluate the global impact of X-ray radiation. \texttt{VPLanet} performs comprehensive simulations of planetary system evolution over Gyr timescale. Its modules cover internal, atmospheric, rotational, orbital, stellar, and galactic dynamics. These modules can be coupled to allow for the simultaneous simulation of the evolution of terrestrial planets, gaseous planets, and stars. For the stellar rotation period evolution, we adopted an approach from \cite{Matt2015}. It is formulated for the torque exerted on a star due to magnetic braking, which depends on the star's mass, radius, rotation rate, and magnetic field strength. The model provides a framework for predicting stellar rotation periods across stellar masses and ages. It also accounts for the dependence of spin-down rates on stellar mass. Lower-mass stars (e.g., M dwarfs) have longer convective turnover timescales. In this study, we employ various modules to calculate the impact of the reported XUV flux from \texttt{VPLanet}, including the \texttt{AtmEsc}, \texttt{STELLAR}, and \texttt{FLARE} modules for simulating atmospheric escape. Atmospheric escape rates depend on a planet’s mass, radius, composition, magnetic field, and orbital distance, while stellar spin-down variability also plays a significant role in cumulative mass loss \citep{Watson1981, Matt_2012,Cohen_2014,Ketzer_Poppenhaeger2023}. Our initial parameters for the modeling are described in Table \ref{tab:vplanet_param}. For the planetary structure model, we follow the study of \citet{Lehmer_2017}, which shows that hydrodynamic escape during early stellar XUV saturation can explain the transition from gas-enveloped to rocky planets, accounting for the observed radius gap in exoplanet populations. The \texttt{AtmEsc} module simulates the escape of planetary atmospheres and the discharge of surface volatiles using energy- and diffusion-limited mechanisms. This module focuses on hydrogen-dominated atmospheres and water vapor-dominated atmospheres. In hydrogen-rich cases, hydrogen escapes first due to diffusive separation, delaying the loss of heavier volatiles. For water vapor atmospheres, the module simulates photolysis, with hydrogen escaping hydrodynamically and dragging oxygen along. 

\begin{table}[h!]
    \centering
    \caption{Parameter of star and planets used for the simulation with \texttt{VPLanet} for K2-18 and K2-18b}
    \begin{tabular}{lc}
    \hline\hline
         Parameter &  Value  \\
         \hline
         Stellar Mass ($M_{\astrosun}$) & 0.413 \\
         Initial Rotation Period (days)  & 1.0\\
         Planet's Mass ($M_{\earth}$) & 8.0\\
         Planet Mass-Radius Model & Lehmer17* / NONE \\
         Planet's Radius (present-day)  ($R_{\earth}$) & 2.38\\
         Planet's Eccentricity &  0.2 \\
         Planet's Semi-major axis (AU)  & 0.143 \\

        Envelope mass ($M_{\earth}$) & 0.06   \\
        Thermosphere temperature (K)  & 400   \\
        Saturated XUV luminosity fraction & $\mathrm{10^{-3}}$ \\
        Initial Age (Myr) & 5.0  \\
        Flare Energy (erg) & $\mathrm{10^{33}}$ to $\mathrm{10^{36}}$ \\
        
    \hline
    \end{tabular}
    \begin{tablenotes}
    \small
      \item  {Reference *\cite{Lehmer_2017} }
    \end{tablenotes}
    \label{tab:vplanet_param}
\end{table}

The \texttt{STELLAR} module in \texttt{VPLanet} models key properties of low-mass stars ($\mathrm{M_{\star} \leq 1.4 M_{\astrosun}}$), such as rotation rate, bolometric luminosity, XUV luminosity, effective temperature, and stellar radius of gyration. In \texttt{VPLanet},  the radius of gyration is used to quantify how a planet's or star's mass is distributed internally, which affects its rotational dynamics. The model relies on \cite{Baraffe2015}. The rotation period and X-ray luminosity relation follows a power law with index -2.26 \citep{Magaudda2020} for non-saturated low mass stars in mass bin $\mathrm{0.4 M_{\astrosun}<M_*<0.6 M_{\astrosun}}$. The model predicts a X-ray luminosity evolution similar to that shown in Figure~\ref{fig:evolutionary_track}, i.e., a factor of several higher than observed at the current epoch. The more sophisticated stellar evolution model with different rotators of CJ+21 and its implied stellar high-energy flux evolution is discussed in Sect. 5. 

The \texttt{FLARE} model \citep{Amaral2022} implemented a power law relationship between flare energy and cumulative flare frequency distribution (number/day) \citep{Davenport2019}, which also depends on the stellar type and age. We note that the \cite{Davenport2019} model overestimates superflare rates, as it is based on younger, more active stars, with only about 3\% of the catalog consisting of M dwarfs \citep{Davenport2016}. This approach allows for a quantitative assessment of the cumulative XUV energy input from random flare events, which complements the quiescent stellar XUV irradiation under the assumption that the flare frequency distribution in EUV has the same power-law index as the one in the optical. The model assumed flare energies between $\mathrm{10^{33}}$ and $\mathrm{10^{36}}$ erg based on \cite{Davenport2019}. The \texttt{FLARE} module computes the average XUV for each simulation time-step, integrating from lowest to highest flare energy. We note that the \cite{Davenport2019} model was developed for younger and more active stars than the Sun, with only 3\% of the catalog stars \citep{Davenport2016} being M-stars. 

\begin{figure*}
    \centering
    \includegraphics[width=\textwidth]{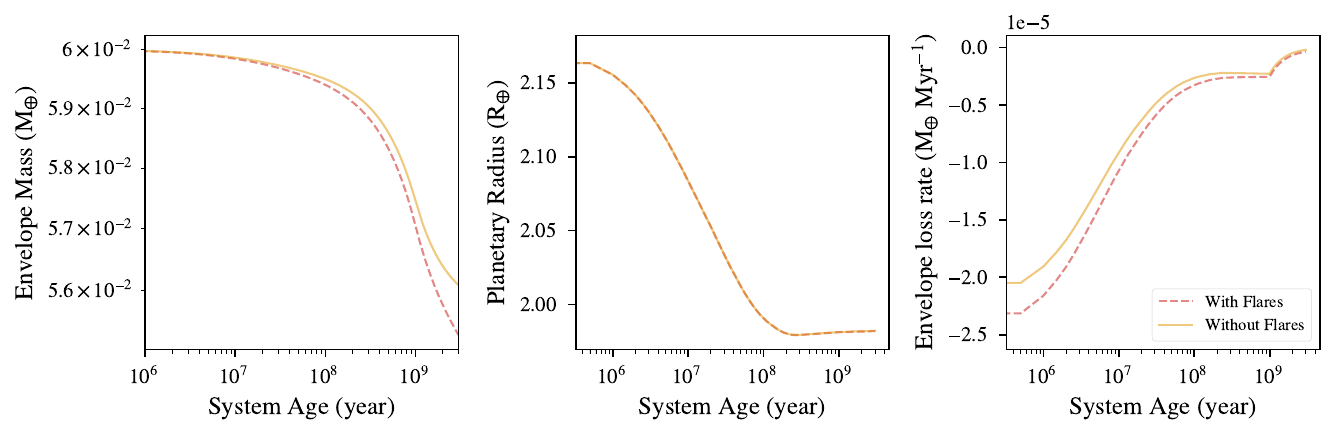}
        \caption{Planetary Evolution of K2-18b. Left panel: Mass of K2-18b over time. The decrease is following the stellar evolutionary track (see Fig.~\ref{fig:evolutionary_track}) in the saturated regime, and assuming the system leaves this regime after 1 Gyr. Middle: the evolution of planetary radius. Right: Envelope loss rate. The red dashed line shows the evolution including the flare contribution \citep{Amaral2022}, while the orange solid line followed the evolution predicted by the \texttt{STELLAR} model \citep{Baraffe2015}. Note that the planet radius shown here is limited by the \cite{Lehmer_2017} model.}
    \label{fig:atm_escape}
\end{figure*}

We model K2-18b as a mini-Neptune, adopting the \cite{Lehmer_2017} framework. The framework provides a model to explain the gap in exoplanet sizes, known as the "radius valley," which separates smaller rocky planets from larger mini-Neptunes with gas envelopes. This model simulates how a planet’s original hydrogen and helium atmosphere is removed by strong X-ray and extreme ultraviolet (XUV) radiation from its host star. The low XUV radiation environment has important implications for the atmospheres of the exoplanets of K2-18. While multiple mechanisms can drive atmospheric mass loss in exoplanets, the only directly observed contributor to date is high-energy (XUV) irradiation. Observational evidence for other drivers, such as stellar winds or magnetic interactions, remains lacking for this system. Therefore, our analysis focuses on hydrodynamic escape processes, including the energy-limited approximation implemented via \texttt{VPLanet} \citep{Barnes2020} for $\mathrm{H_2}$–$\mathrm{H_2O}$-dominated atmospheres (see Fig. \ref{fig:atm_escape}).

In the case of $\mathrm{H_2/H_2O}$-dominated atmospheres, we use the \texttt{VPLanet} model. The Mini-Neptune scenario is implemented using the \texttt{AtmEsc} module, which simulates atmospheric escape processes, reproducing the results from \citet{Lopez_Fortney2013}. It demonstrates how the evolution of a Mini-Neptune's gaseous envelope depends on its core mass, initial envelope mass, and total mass. 

We use the \texttt{AtmEsc} module of \texttt{VPLanet} to provide an overview of atmospheric evolution over time. We note, however, that the energy-limited approach can be quite uncertain and in fact is an order-of-magnitude estimate for the upper limit of the mass-loss rate only, see \cite{Krenn2021} for more details. The result of the model is shown in Fig.~\ref{fig:atm_escape}. The envelope mass evolves by 5\% over 1 Gyr, even when stellar flare contributions are included. This suggests that, despite this evolution with flare contribution, the planet can maintain a significant hydrogen-rich envelope. This resilience supports the potential of habitable conditions (liquid water in the form of, e.g., surface or subsurface oceans) under a stable envelope in some situations, particularly for so-called Hycean or stratified mini-Neptune models. 


\end{appendix}
\end{document}